\documentclass[%
 reprint,
 superscriptaddress,
 amsmath,amssymb,
 prx,
]{revtex4-2}

\usepackage{amsmath, amssymb}
\usepackage{graphicx}
\usepackage{dcolumn}
\usepackage{bm}
\usepackage{bbm}
\usepackage{microtype}
\usepackage{dsfont}
\usepackage{mathtools}
\usepackage{comment}
\usepackage{overpic}
\usepackage{float}
\usepackage{braket}
\usepackage{color} 
\usepackage{soul}
\usepackage[normalem]{ulem}
\usepackage[
    colorlinks  = true, 
    linkcolor   = blue,
    urlcolor    = blue,
    citecolor   = blue,
    anchorcolor = blue]{hyperref}

\newcommand{\Kll}[1]{\textcolor{black}{#1}}

\newcommand{\mbf}[1]{\boldsymbol{#1}}

\begin{document}

\title{Noise-Robust Detection of Quantum Phase Transitions}

 \author{Kevin Lively}%
  \affiliation{%
 Institute for Software Technology, German Aerospace Center (DLR), 51147 Cologne, Germany
}%
 \author{Tim Bode}
  \affiliation{%
Institute for Quantum Computing Analytics (PGI-12), Forschungszentrum Jülich, 52425 Jülich, Germany
}%
 \author{Jochen Szangolies}%
  \affiliation{%
Institute for Software Technology, German Aerospace Center (DLR), 51147 Cologne, Germany
}%
\author{Jian-Xin Zhu}%
\affiliation{Theoretical Division, Los Alamos National Laboratory, Los Alamos, New Mexico 87545, USA}
\affiliation{Center for Integrated Nanotechnologies, Los Alamos National Laboratory, Los Alamos, New Mexico 87545, USA}

\author{Benedikt Fauseweh}%
 \affiliation{%
Department of Physics,  TU Dortmund University, Otto-Hahn-Str 4, 44227 Dortmund, Germany
}%
  \affiliation{%
 Institute for Software Technology, German Aerospace Center (DLR), 51147 Cologne, Germany
}%

\date{\today}
             
\begin{abstract}
Quantum computing allows for the manipulation of highly correlated states whose properties quickly go beyond the capacity of any classical method to calculate. Thus one natural problem which could lend itself to quantum advantage is the study of ground-states of condensed matter models, and the transitions between them. However, current levels of hardware noise can require extensive application of error-mitigation techniques to achieve reliable computations. In this work, we use several IBM devices to explore a finite-size spin model with multiple `phase-like' regions characterized by distinct ground-state configurations. Using pre-optimized Variational Quantum Eigensolver (VQE) solutions, we demonstrate that in contrast to calculating the energy, where zero-noise extrapolation is required in order to obtain qualitatively accurate yet still unreliable results, calculations of the energy derivative, two-site spin correlation functions, and the fidelity susceptibility yield accurate behavior across multiple regions, even with minimal or no application of error-mitigation approaches. Taken together, these sets of observables could be used to identify level crossings in a simple, noise-robust manner which is agnostic to the method of ground state preparation. This work shows promising potential for near-term application to identifying quantum phase transitions, including avoided crossings and non-adiabatic conical intersections in electronic structure calculations.
\end{abstract}

\maketitle

\section{Introduction}
The promise of quantum computing resides in making classically infeasible computations realizable. The simulation of quantum systems, known to be a classically hard task, yields a particularly natural use-case.
However, before the advent of scalable, fault-tolerant quantum computers, extracting relevant results from the currently available `noisy intermediate-scale quantum' (NISQ) devices remains a challenge due to high levels of noise from state preparation, measurement, control, cross talk, etc. These lead to errors which can seriously degrade the results of computations using the experimentally measured data, and vary over time in non-trivial ways \cite{Dahlhauser2021,Fauseweh2021,dasgupta2023reliability,Kim2023,Thorbeck2023}. Despite these issues, assertions of the advantage of quantum computation on NISQ devices over classical methods have been made, focused on highly artificial problems that have no immediate real-world application \cite{Arute2019, Zhong2020, Wu2021, Zhu2022}. Furthermore, this preliminary evidence for useful quantum computation before the realization of fully error-corrected quantum devices remains a contentious topic, with many claims being challenged via refinements of existing classical computational schemes \cite{beguvsic2023fast,tindall2023efficient,kechedzhi2024effective, Pan2022}. 

Consequently, a conclusive real-world use case experimentally demonstrating quantum advantage has yet to be identified. Given that in their most naive interpretation, Quantum Computers are devices capable of preparing and manipulating high dimensional states, a natural use case for them is the investigation of the highly entangled and classically intractable ground and dynamical states of correlated condensed matter systems, which can yield novel insights to material structure, function and non-equilibrium properties \cite{Fauseweh2023quantumcomputing,Fauseweh2024,PhysRevResearch.6.033092}.

In this pursuit to utilize the capabilities of near- to medium-term available quantum devices for such practically significant problems, recent years have seen the development of a large number of techniques to leverage NISQ hardware. Broadly speaking these can be thought of as algorithmic developments and methods of Error Mitigation (EM) which partially compensate for the errors present in computation via pre- and post-processing to bring expectation values of observables closer to their ideal noise-free levels. The topic of EM is a rapidly evolving field: for a thorough recent review of the state of the art, see \cite{Cai2023}. On the algorithmic side, one major area of investigation is that of ground state preparation for arbitrary Hamiltonians. Methods such as adiabatic state preparation, quantum phase estimation, Imaginary Time Evolution and the Dissipative Quantum Eigensolver are all active areas of research, for an overview see \cite{cubitt2023, Nishi2023} and references therein. 

Variational Quantum Algorithms (VQAs) based on Parameterized Quantum Circuits (PQCs) in particular have garnered a lot of interest on theoretical and pragmatic grounds \cite{Cerezo2021,Tilly2022}. On the theoretical side the expressive power of PQCs has been shown to outperform generative neural networks \cite{Du2020}, thereby surpassing Neural Quantum States, which themselves can exceed limitations of Tensor Network states widely considered to be a gold standard for area-law entangled states \cite{lange2024}. Pragmatically these algorithms are straightforward to implement on current and near-term devices, and it is primarily for this reason that we utilize this method here.

In brief, a VQA is a hybrid quantum-classical algorithm that utilizes classical optimization techniques to iteratively refine a parameterized ansatz quantum circuit aimed at, e.g., preparing some target quantum state or minimizing a target cost function. In particular, the Variational Quantum Eigensolver (VQE, \cite{Peruzzo2014}) aims to find the ground state of a quantum system by minimizing the quantity
\begin{equation}
    E(p)=\braket{\Psi(p)|\hat{H}|\Psi(p)}
\end{equation}
for a given Hamiltonian $\hat{H}$, where $\ket{\Psi(p)} = \hat{C}(p)\ket{\Psi_0}$ is varied by means of adjusting the parameters $p$ defining the PQC $\hat{C}(p)$ acting on the initial state $\ket{\Psi_0}$. This yields an upper bound to the ground-state energy $E_G$, attaining the actual value if and only if $\ket{\Psi(p)}$ is the system's ground state. 

Regardless of the method by which one prepares the system ground state, the capacity to directly manipulate these states could have a significant impact in one of the most important challenges in condensed matter physics: the characterization of phase diagrams, particularly quantum phase transitions occurring at $T=0$ which are defined by the rearrangement of ground states driven by variations of Hamiltonian parameters. These problems can be extremely challenging for analytical or classical numerical methods to handle due to the quantum critical points (QCPs) defining the transition being characterized by highly entangled states \cite{Hauke2016}, in particular for highly frustrated systems with many competing phases \cite{Dutta2015}. 

Subsequently developing methods to identify phases of matter using NISQ hardware and algorithms has received significant attention in recent years. While explorations of equilibrium transitions have seen the application of a plethora of methods including calculating correlation functions in different phases \cite{Borzenkova2021,Scholl2021,Ebadi2021}, non-local string order parameters and Chern numbers for topological phases \cite{Smith2022,Xiao2023}, tracking changes in fidelity of ground states between Hamiltonian parameters \cite{Okada2023}, entanglement spectrum \cite{Xiao2021}, \Kll{coupling systems to external probes which act as witness to transitions \cite{Wang2007,Ai2009},} machine learning assisted 
 variational optimization \cite{cao2024} or classification by Quantum Convolutional Neural Networks \cite{Liu2023,zapletal2023errortolerant,Herrmann2022}, the majority of these studies were either purely classical simulations, digital quantum simulations with ground states prepared by hand from known or classically solved instances, or analogue experiments with limited gate sets. Furthermore those studies which used digitally programmable quantum hardware were generally restricted to a small number of physical qubits, typically less than five, or with limited EM techniques applied.

In contrast, the goal of this paper is to demonstrate that determining phase transitions from states prepared on NISQ devices requires the effective utilization of a collection of noise robust observables working in tandem. To accomplish this we apply the VQE to investigate ground-state properties of a frustrated spin model whose ground state manifold displays a rich variety of highly entangled correlated states, and utilize real hardware to explore its properties for system sizes up to twelve sites. Utilizing IBM superconducting transmon devices, we particularly focus on the question of which observables can be reliably calculated despite noise in order to determine, potentially without prior knowledge, where transitions occur from these properties alone. However, we find that the noise level of these devices is still prohibitively high when investigating a simple scalar quantity characterizing the system, such as the ground-state energy, failing to even reproduce the qualitative behavior when using parameters pre-optimized via classical simulation. Using EM techniques ameliorates the problem somewhat, but ultimately proves insufficient to produce reliable results or to extract nontrivial features of the system's phenomenology. Thus, while a narrow focus on the energy to characterize the system seems to show strong limitations for the conclusions available using current hardware, expanding our scope to more noise-robust quantities suggests that this is merely due to an inefficient utilization of data.

Instead, we propose to shift focus to features whose qualitative behavior might be recovered even within NISQ limitations. By exploring several different observables for each VQE solution, we demonstrate that utilizing the measurement data already gathered for the energy calculation in order to construct the first energy derivative and spin-spin correlation functions results in very clean experimental signals of ground state rearrangement under different levels of EM. Furthermore we show that appropriately adapting the Fidelity Susceptibility (FS) to the potentially symmetry breaking expressivity of the VQE ansatz yields a direct measurement of the similarity of different VQE solutions with a high signal to noise ratio, which can be used in the case of topologically differentiated states where local order parameters would fail. We therefore argue that by cross referencing these measurements, one is more capable of accurately inferring whether large fluctuations in a given subset of data are due to corruption from noise or stem from really existing `hidden' noise-free properties of the encoded wavefunction. Consequently, we find that when working within the limitations of present-day quantum devices, while noise may wash out even qualitative features of observables such as energy, useful information can still be obtained by focusing on noise-robust quantities such as the FS, alongside other observables, and therefore that the choice of observable may be just as important as EM if one wants to unambiguously demonstrate a meaningful result beyond the limitations of classical computation.

The rest of the paper is structured as follows. In Sec.~\ref{sec:ANNNI}, we describe our frustrated spin model in more detail, and discuss the method of studying phase transitions using FS. Sec.~\ref{sec:Experiment} introduces our experimental methodology, discussing the ansatz used for our VQE-circuit and the EM methods used. The following Sec.~\ref{sec:Energy} showcases the results of experiments performed on IBM quantum computers to calculate the energy, and demonstrates the difference between using `raw' data and EM experiments. Sec. ~\ref{sec:CorrFid} shows that the energy derivative, coupled with correlation functions and with FS measurements displays unambiguous signals of ground-state rearrangement; we demonstrate that the latter two observables display noise robust behavior across several different choices of system size, boundary conditions and usages of EM. We conclude in Sec.~\ref{sec:Conclusion} with a discussion of the implications of our findings for the overriding goal of finding quantum utility by suggesting several applications of our method to specific systems.

\section{Axial Next-Nearest Neighbor Ising (ANNNI) Model}\label{sec:ANNNI}

\begin{figure}[htb!]
    \centering
    \vspace{0.5mm}
    \includegraphics[width=\linewidth]{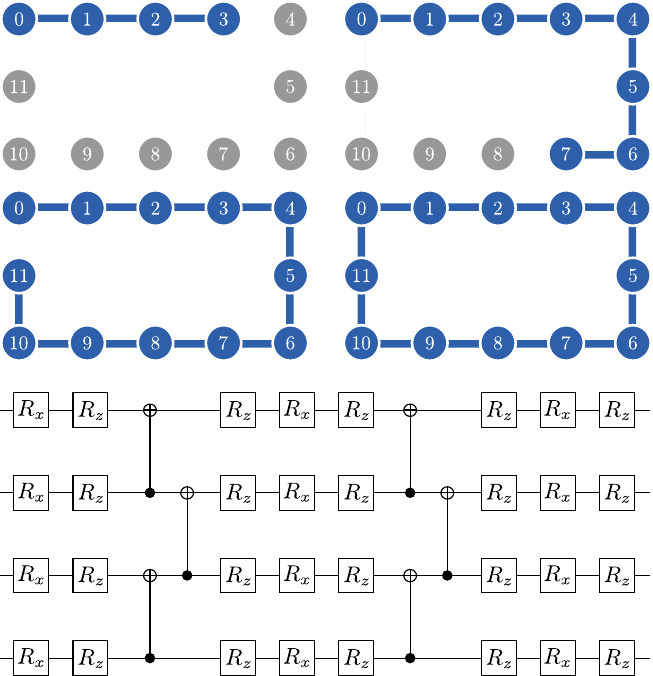}
    \caption{ The top four plots show example circuit layouts of the VQE ansatz mapped to the IBM heavy hexagon architecture, showing $N=4$ and $N=8$ under open boundary conditions on top, and $N=12$ with open and periodic boundary conditions respectively below. The structure of one layer of the VQE ansatz for $N=4$ is shown on the bottom. For $N=8$ and $12$ the structure is effectively the same, but tiled vertically.}
    \label{fig:VQE-ansatz}
\end{figure}

Throughout this work we study the ground-state properties of VQE solutions to the ANNNI model \cite{Selke1988}, written in terms of the Pauli operators on qubits $i\in [0,N-1]$, $Z_i, X_i$, as
\begin{equation}\label{eq: hamiltonian}
     \hat H(p) = -J_1\sum_i Z_iZ_{i+1} + J_2\sum_iZ_iZ_{i+2} + B_x\sum_iX_i,
\end{equation}
where $\hat H$ is parameterized by $p=(J_2, B_x)$. We consider $J_2> 0$, set $J_1=1$ as the unit of energy and use both open and periodic boundary conditions. Despite its seeming simplicity, the combination of a nearest-neighbor ferromagnetic Ising coupling with a \emph{next}-nearest neighbor \emph{anti}-ferromagnetic coupling alongside fluctuations induced by the transverse field $B_x$, playing a role analagous to temperature, induces a rich variety of phases, including modulated magnetic orders, commensurate to incommensurate transitions, floating phases, and potential Koster-Thouless infinite order transitions \cite{Bak1982,Selke1988,Arizmendi1991,Suzuki2012,Dutta2015,Bonfim2017,Fumani21}. Although one must of course be in the thermodynamic limit in order to speak of true phases, the structure of these ground states is distinct enough to allow one to identify `phase-like' regions. Thus we use the terms `critical point', `phases' and `transitions' when referencing these different regions of parameter space defined by the structure of the ground-state wavefunction.

Since ANNNI essentially encapsulates the complexity of correlated matter through competition and fluctuation, its phase diagram is a subject of continuing interest in the literature. Yet, there remains no consensus on the number, nature, or location of the phases present \cite{Bonfim2017,cea2024}, despite investigation through a prolific set of methods, including quantum convolutional neural networks \cite{Monaco2023,cea2024}, matrix product states \cite{Nagy2011}, density-matrix renormalization group (DMRG) \cite{Beccaria2006}, quantum monte carlo \cite{Arizmendi1991} and perturbative approaches \cite{Chandra2007}. \Kll{For example, Beccaria et al. \cite{Beccaria2006} and Nagy \cite{Nagy2011} both use Matrix Product State based methods and report that all phase transitions converge at one multi-critical point. By contrast the perturbative approach of Chandra and Dasgupta \cite{Chandra2007} finds that the paramagnetic phase is restricted to sufficiently high $B_x$.} 

One particularly challenging phase to describe is the floating phase, suitable for describing the properties of the frustrated magnetic compound Ca$_3$Co$_2$O$_6$ \cite{Batista2012,Zapf2022} which displays characteristics indicative of incommensurate phases formed of periodic order parameters mismatched with respect to the underlying lattice constant \cite{Bak1982}. While there are very strong theoretical tools for describing gapped phases in 1D and 2D, if the dominant correlation displays oscillations of this kind, requiring large unit cells to capture, it becomes challenging for 1D methods \cite{Beccaria2006,Mila2022} and virtually impossible for 2D systems, while still being experimentally accessible \cite{zhang2024probing}. \Kll{In the case of the ANNNI model, the study of Beccaria et al. reports that the power law dictating the correlation function which characterizes the phase converges to two different possible exponents depending on the number of sites, and thus even the gold standard of 1D models, DMRG, breaks down.} Therefore given the potential for NISQ devices to access highly such correlated states, the model forms an ideal testbed for promising approaches to digital quantum simulation by combining a simple system with a well-understood, non-trivial phenomenology.

\Kll{The aspects of the phase diagram for this model which are broadly agreed on are that} for $B_x=0$, it is trivial to show that there are two ground-state regimes divided by the critical point $J_2^c = 1/2$. For $J_2<1/2$ the ground state adopts an energy-degenerate ferromagnetic superposition, namely $\ket{\Psi} \approx \frac{1}{\sqrt{2}}(\ket{\uparrow}^{\otimes N} + \ket{\downarrow}^{\otimes N})$. Whereas for $J_2>1/2$ the next nearest neighbor anti-ferromagnetic term dominates and the ground state adopts what is referred to as a $\braket{2,2}$ configuration, characterized by spin states looking like $\ket{\uparrow\uparrow\downarrow\downarrow\uparrow\uparrow\downarrow\downarrow\ldots}$. The ground state at the critical point $(J_2^c,0)$ is dominated by fluctuations and is completely degenerate, but numerical studies have shown that lifting the degeneracy by setting $B_x\neq 0$, reveals this to be a multi-critical point of a number of phases dependent on boundary conditions and system size $N$, fanning out as $B_x$ increases until reaching a paramagnetic phase for $B_x\gg J_2$. 

This rich diversity of phases is characterized by phase-transitions which are generally understood to be second order in nature, with the exception perhaps of the phase transition to the $\braket{2,2}$ phase, which could first order \cite{Bonfim2017}. Thus identifying these transitions requires utilizing either an appropriately constructed order parameter, or investigation with correlation functions. As already mentioned, for incommensurate or floating phases this can become a significant hurdle. An alternative metric which has been used to detect ground-state configuration rearrangements in condensed matter and universal properties of critical points in an order-parameter agnostic manner is the Fidelity Susceptibility. Given any circuit which prepares a ground state, one can directly calculate this quantity on a Quantum Computer. However, when dealing the with VQE specifically, we must adapt this method to partially compensate for its shortcomings.

\subsection{Fidelity Susceptibility}\label{sec:FidSus}
Given a wavefunction dependent on a parameter $p$ (restricted to be one dimensional for simplicity) the sensitivity of the fidelity of the wavefunction with respect to a change in this parameter $\delta$ is given by 
\begin{equation}\label{eq: fidelity macro delta}
\begin{split}    
    F(p,\delta) &= |\braket{\Psi(p - \delta)|\Psi(p + \delta)}|.
\end{split}
\end{equation}
Taking the Taylor expansion of this expression to second order, one obtains
\begin{equation}\label{eq: fidelity expansion in delta}
\begin{split}    
    F(p,\delta) &= 1-\chi(p)\delta^2 + \mathcal{O}(\delta^4),
\end{split}
\end{equation}
which defines the Fidelity Susceptibility (FS) $\chi(p)$. This measure has been used as a standard tool in classical analysis to study phase transitions in condensed matter systems \cite{Bonfim2017,Wang2015,Rossini2021}. It is conceptually related to the Loschmidt echo and has direct bearing on the Quantum Fisher Information \cite{Lambert2023}. Furthermore, it is a measurable quantity via experiments in the linear response regime \cite{Gu2014} and has been studied directly on NMR devices \cite{Zhang2008,Zhang2009}. Throughout this work we use the pragmatic approximation $\chi(p)\approx 1-|\braket{\Psi(p)|\Psi(p + \delta)}|$ for ease of calculation. For each point of interest in the phase space $p$, one can optimize a VQE ansatz parameter vector, $\mbf{\phi}(p)$ such that $E_{\text{VQE}}(\mbf{\phi}(p)) \approx E_{\text{G}}(p)$, and subsequently calculate $\chi$ between different solutions $\ket{\Psi_{\text{VQE}}(\mbf{\phi}(p))}$ (which for ease of notation we define to be $\ket{\Psi_{\text{VQE}}(p)} = \hat{C}(\mbf{\phi}(p))\ket{0}$) via measuring the proportion of outcomes in the $\ket{0}$ state following preparation of $\hat{C}^{\dagger}(\mbf{\phi}(p'))\hat{C}(\mbf{\phi}(p))\ket{0}$ \cite{Okada2023}.

There is however an important caveat to calculating the FS between VQE solutions, without explicitly turning to differentiation with respect to the VQE parameters \cite{Matteo2022}. Given that the expressibility of any given circuit ansatz is limited, it's not guaranteed that a given solution $\ket{\Psi_{\text{VQE}}(p)}$ preserves the symmetries of the system if the optimization manifold biases a particular broken symmetry or the solution is trapped in a local minimum. If two nearby VQE solutions should happen to break the system symmetries in non-commensurate ways, the fidelity between these states will be artificially suppressed. However, this effect can be partially countered.

Every symmetry operation $i$ of the Hamiltonian can be represented by a unitary operator $\hat{U}_i$ such that  $[\hat{H},\hat{U}_i] = 0$. We can collect these operators into the group $\mathcal{U}$. By definition a symmetry-broken solution will not be invariant under some $\hat{U}_i\in\mathcal{U}$, i.e. $|\braket{\Psi_{\text{VQE}}(p)|\hat{U}_i|\Psi_{\text{VQE}}(p)}|< 1$. The state $\hat{U}_i\ket{\Psi_{\text{VQE}}(p)}$ instead corresponds to a rotation within the energy degenerate subspace dictated by symmetry operation $i$. We can generalize this notion by using $\hat{U}_i$ as a generator of rotations characterized by $\hat{R}_i(\theta) = \exp\left(i\theta\hat{U}_i\right)$. If $\hat{U}_i$ is not Hermitian, linear combinations of $\hat{U}_i$ within the same symmetry class can be performed to create Hermitian operators so that the operator exponential is guaranteed to preserve state normalization. We denote all possible rotations associated with each generator in the group as
\begin{equation}\label{eq: symmetry rotations}
    \hat{R}(\vec{\theta}) = \exp\left(i\sum_i\theta_i\hat{U}_i\right),
\end{equation}
where $\vec{\theta}\in\left[0,2\pi\right]^{d-1}$ where $d=|\mathcal{U}|$ is the cardinality of $\mathcal{U}$, and we ignore the trivial identity operation $\hat{I}$ corresponding to the $U(1)$ global phase symmetry.

 With these definitions in mind we can consider the VQE solutions at two phase space points $p$ and $p'$. We must try to align these two solutions as much as possible within the degenerate subspace in order to approximate the fidelity measure. Thus the calcuation becomes:
\begin{equation}\label{eq: Rotated Fidelity}
    F_{\text{VQE}}(p,p') = 1-\max_{\vec{\theta}}|\braket{\Psi_{\text{VQE}}(p)|\hat{R}(\vec{\theta})|\Psi_{\text{VQE}}(p')}|.
\end{equation}
Such rotations are not guaranteed to be capable of bringing arbitrary superpositions into perfect alignment with each other. Indeed it is enough to see this by comparing two solutions at the same point. Consider one VQE parameterized solution $\ket{\Psi_{\text{VQE}}(p)}$ which is in an equal superposition within the degenerate subspace. By construction this state is invariant under all symmetry operations such that $|\langle \Psi_{\text{VQE}}(p)| \hat{U}_i|\Psi_{\text{VQE}}(p)\rangle| = 1$ for all $\hat{U}_i\in \mathcal{U}$. Now consider another solution $\ket{\tilde{\Psi}_{\text{VQE}}(p)}$ in a completely symmetry broken state, which by construction means that $|\langle \tilde{\Psi}_{\text{VQE}}(p)| \hat{U}_i|\tilde{\Psi}_{\text{VQE}}(p)\rangle| < 1$ for all $\hat{U}_i\neq \hat{I}$. A concrete example for a ferromagnetic ground state with respect to the $\mathbb{Z}_2$ symmetry under open boundary conditions would be $\ket{\Psi} = \frac{1}{\sqrt{2}}(\ket{\uparrow\uparrow\ldots} + \ket{\downarrow\downarrow\ldots})$, and $\ket{\tilde{\Psi}} = \ket{\uparrow\uparrow\ldots}$. In this case one could represent the $\mathbb{Z}_2$ symmetry as $\hat{U} = X^{\otimes N}$, and $\mathcal{U} = \{I, X^{\otimes N}\}$. It is easy to see that the equal superposition state is, up to a phase, unaffected by the rotations in Eq.~(\ref{eq: Rotated Fidelity}) and thus $|\braket{\tilde{\Psi}|\hat{R}(\vec{\theta})|\Psi}|$ effectively measures the projection of the $d=2$ dimensional superposition onto one basis state in the symmetry subspace, corresponding to $1/\sqrt{d}$. 

Depending on the number of symmetries present, scanning over the $d$ dimensional sphere of rotation angles can become prohibitively expensive, even when doing coarse sampling. Furthermore, implementation of Eq.~\ref{eq: symmetry rotations} can lead to a large experimental overhead depending on the native gate set available. Thus a pragmatic approximation which we use throughout this work is to take only the effect of the generators:
\begin{equation}\label{eq: practical fidelity}
    F_{\text{VQE}}(p,p') \approx 1-\max_{i}|\braket{\Psi_{\text{VQE}}(p)|\hat{U}_i|\Psi_{\text{VQE}}(p')}|.
\end{equation}
In our case, we precompute the appropriate generators to apply between states across the parameter scan, consisting of total bit flip, $X^{\otimes N}$, and when working in periodic boundary conditions, the shift operators $L_i$ which send site $i\to (i+k)\ \text{mod}\ N$. We implement shift operators by simply reassigning qubits to the inverted parameterized VQE unitary used to calculate the overlap between states.

\section{Experiments}\label{sec:Experiment}
\subsection{Variational Circuit and Error Mitigation}
The structure of the VQE ansatz used in this study is epitomized in Fig.~\ref{fig:VQE-ansatz}, although this figure is given for $N=4$ sites for clarity. Each ansatz layer consists of two rounds of CNOT blocks applied to the qubits in a staggered fashion, minimizing idle time. We choose to construct each layer with two rounds of entangling gates in order to induce entanglement between next nearest neighbors within the connectivity restrictions of the heavy-hex topology also depicted in Fig.~\ref{fig:VQE-ansatz}. All the results presented are for a single optimization layer. The optimal parameters for the ANNNI model were converged using a classical gradient descent simulation as implemented in the YAO package \cite{Luo2020}. For the $N=12$ VQE, in order to improve convergence the initial cost function was chosen so as to maximize overlap with the classically computed exact ground state. However, we removed any underlying symmetries in the ground state, e.g. $\mathbb{Z}_2$ inversion, in order to reduce the degree of superposition in the target state. Following convergence, the cost function was switched back to the energy, and the previously converged parameters used as a starting guess, which tended to reintroduce some superposition across the symmetry subspace. For $N=8$ and $4$ we simply used the energy cost function throughout the optimization.

In order to mitigate readout error we utilized Twirled Readout Error eXtinction (TREX)\cite{vanDenBerg2022} which effectively diagonalizes the readout error map through random application of Pauli strings in $\{I,X\}^{\otimes N}$ to the qubits, followed by inversion in post processing, at the cost of running one set of calibration circuits. We performed TREX calibration measurements at each $J_2$ value in the scan before measuring in the $X$ and $Z$ bases and performing the FS calculation. We furthermore utilized Zero Noise Extrapolation \cite{Temme2017} by replacing every instance of a CNOT gate with three and five physically applied, but logically redundant CNOT gates, under the expectation that the noise associated with two qubit gates is the dominant source of error. We used an exponential noise model $E(\lambda) = E_0\exp\left(a\lambda\right)$ to fit $E_0$ and $a$ for $\lambda \in \{1,3,5\}$. In order to eliminate coherent error, we utilized Pauli Twirling of our CNOT gates, which has been shown experimentally to enforce a stochastic Pauli noise channel associated with a significantly reduced worst-case error rate \cite{Wallman2016,Ware2021}. When performing TREX or Twirling we use 16 circuit instantiations, which has been credited as being sufficient for many applications \cite{vanDenBerg2023,Kim2023} and used 100,000 shots distributed across the 16 circuits. We implemented these tools by hand in Qiskit using the QiskitRuntime Sampler primitive \cite{Qiskit}. 

\subsection{Energy}\label{sec:Energy}
\begin{figure}
    \centering
    \includegraphics[width=\linewidth]{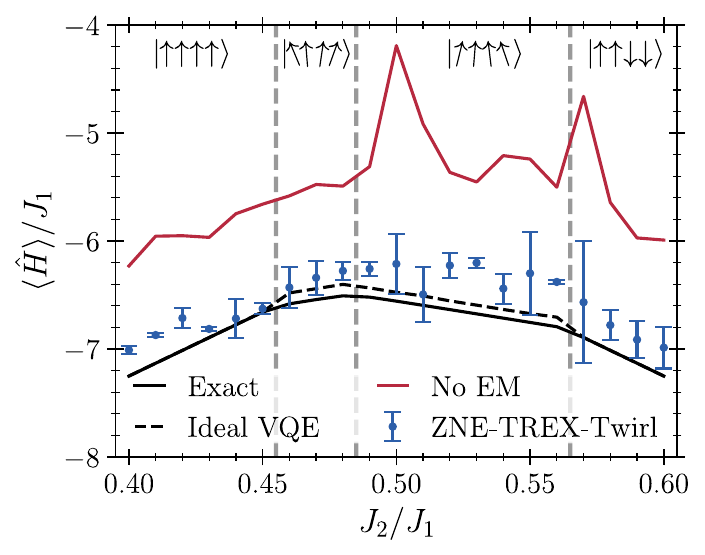}
    \vspace{-0.5cm}
    \caption{The ground-state energy profile of Eq.~\ref{eq: hamiltonian} for $N=12$ at $B_x=0.1$ under periodic boundary conditions obtained via executing the pre-optimized VQE solutions on \texttt{ibm\_osaka} with no EM in red and \texttt{ibm\_kyoto} with EM in blue. Error bars are due to shot noise in the former case (invisible on this scale) and ZNE fitting and extrapolation uncertainty in the latter. The exact diagonalization solution is in solid black with the ideal, noise free, VQE values in dashed black obtained via a statevector simulation. Cartoons of the ferromagnetic, floating and $\braket{2,2}$ phases are indicated at the top of the graph with vertical grey dashed lines at the exact transition points, determined by $\chi$.}
    \label{fig:ZNE-Energy}
\end{figure}
In Fig.~\ref{fig:ZNE-Energy}, we show indicative experimental results, obtained by running the precomputed optimal VQE parameters for the ground-state energy $E$ of Eq.~(\ref{eq: hamiltonian}) for different $J_2$ values across multiple devices at $B_x=0.1$ for $N=12$ under periodic boundary conditions. Both of these experiments were performed on 127 qubit `Eagle r3' IBM machines. The noise-free statevector simulation of the optimized parameters in dashed black is relatively well converged to the numerically exact solution obtained via exact diagonalization, in particular in the `more classical' ferromagnetic and floating phases, in which the degree of superposition in the Z-basis is minimized compared to the $B_x$ term dominated floating phases. 

The results from experiments on real devices without error mitigation is quite poor, with consistently high energy values and spurious deviations from the underlying inverse U trend of the ideal energy profile. When utilizing TREX, Twirling and ZNE in concert, we can obtain somewhat more reasonable results, with smaller spurious deviations from the ideal values, and a qualitatively acceptable profile. Without using TREX and twirling, the ZNE fits have large uncertainties, leading to larger error bars and unphysical extrapolated energy values for some runs. See Fig.~\ref{fig:CX scaling} in the Appendix for more details. 

\subsection{Noise Robust Detection of Phase Changes}\label{sec:CorrFid}
Given the large extrapolation uncertainty in some of the ZNE points, alongside the potential to return non-physical extrapolations, it's hard to claim without prior knowledge that any given energy calculation is useful. Can we instead find a better use for the same experimental results? Naturally, in order to calculate the energy, measurements must be done in the $X$ and $Z$ bases, and the energy calculated via the sums present in the Hamiltonian, Eq.~(\ref{eq: hamiltonian}). This single scalar is of course far from all the information which can be calculated from a measurement in these bases. We can, \textit{from the same experimental data}, easily calculate other quantities. 

In order to learn more about the properties of the system, and in particular to study the `phase transitions' which occur, we start by looking at the derivative of the energy with respect to $J_2$ through the parameter space. While analytical expressions for higher order derivatives of the energy within the VQE have seen substantial development in recent years \cite{Mitarai2020,kuroiwa2022quantum,Yalouz2022,OBrien2022,Omiya2022,ahmed2022implicit}, due to their relevance in quantum chemistry applications, we consider simply the first order derivative of the ground state energy. Assuming that the VQE solution is close enough to the exact (expressible) ground state, this can be straightforwardly calculated via the Hellmann-Feynman force \cite{kuroiwa2022quantum,Mitarai2020}: $\partial_{J_2}E = \braket{\Psi(p)|\partial_{J_2}\hat{H}|\Psi(p)}$, which clearly entails taking only the expectation value of the anti-ferromagnetic sum in Eq.~(\ref{eq: hamiltonian}), referred to here as $\hat{H}_{A}$. It turns out that this quantity can serve a similar role as the FS in determining phase transitions, which we briefly clarify here. We start by taking the second derivative of the energy: $\partial_{J_2}^2E = \braket{\Psi(p)|\hat{H}_{A}|\partial_{J_2}\Psi(p)} + c.c$. From the first order perturbative expansion of $\ket{\partial_{J_2}\Psi(p)}$ (for the exact state) we have:
\begin{equation}
    \ket{\partial_{J_2}\Psi(p)} = \sum_{n\neq 0} \frac{H_{A}^{n0}\ket{\Psi_n(p)}}{E_n(p)-E_0(p)},
\end{equation}
where $H_{A}^{n0} = \braket{\Psi_n(p)|\hat{H}_{A}|\Psi_0(p)}$ for the excited states $\hat{H}(p)\ket{\Psi_n(p)} = E_n(p)\ket{\Psi_n(p)}$ with energies $E_n(p)$. Subsequently:
\begin{equation}\label{eq: second order energy derivative}
    \partial_{J_2}^2E(p) = 2\sum_{n\neq 0}\frac{|H_{A}^{n0}|^2}{E_n(p)-E_0(p)},
\end{equation}
which is seen to be the second order perturbative correction to the energy with respect to $J_2$. A similar perturbative analysis of equations (\ref{eq: fidelity macro delta}) and (\ref{eq: fidelity expansion in delta}), straightforwardly leads to an expression for $\chi$ in the limit that $\delta\to 0$ \cite{Wang2015,Rossini2021}:
\begin{equation}\label{eq: perturbative chi}
    \chi(p) = \sum_{n\neq 0} \frac{|H_A^{n0}|^2}{(E_n(p)-E_0(p))^2}
\end{equation}
Clearly Eqs. (\ref{eq: perturbative chi}) and (\ref{eq: second order energy derivative}) have a very similar form up to the exponent in the denominator. Thus we can expect that when the gap closes at a phase transition that both measures will diverge, although there will be differences depending on if there is a first or second order phase transition present when in the thermodynamic limit, the Anderson orthogonality catastrophe notwithstanding \cite{Anderson1967}. Since it is trivial to plot the first energy derivative, we can inspect this data for large changes as evidence of a strong second derivative, and thus evidence for a phase transition or level crossing, again with the caveat that we are working with finite sized systems.

\begin{figure}
    \centering
    \includegraphics[width=\linewidth]{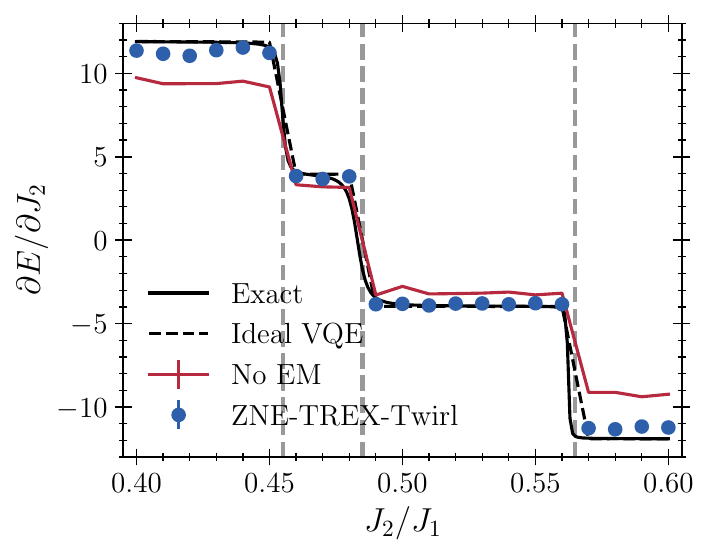}
    \vspace{-0.5cm}
    \caption{The first derivative of the energy for $N=12$ at $B_x=0.1$ under periodic boundary conditions, in units of $J_1$ obtained via executing the pre-optimized VQE solutions on \texttt{ibm\_osaka} with no EM in red and \texttt{ibm\_kyoto} with EM in blue. Exact diagonalization results are in solid black, while the noise free statevector simulation of the VQE is in dashed black. The grey dashed vertical lines indicate the phase transition points determined by $\chi$.}
    \label{fig:diabat energies}
\end{figure}

In Fig.~\ref{fig:diabat energies} we show the results of calculating the first energy derivative for the same data set seen in Fig.~\ref{fig:ZNE-Energy}. It is obvious at a glance that the experimental data clearly recovers the step transitions between the different phases, even when performing no EM. Nonetheless the suite of EM considered here considerably improves the quantitative accuracy of the results, and remarkably the ZNE fits have extremely little variance leading to quite tight error bars, not visible on this scale. This is potentially due to the entire measurement data coming from a single set of $Z$ basis measurements for different $\lambda$ CX levels performed sequentially (thus with minimal drift in noise channels due to non-Markovianity), and being used to calculate just one antiferromagnetic sum, as opposed to separate $X$ and $Z$ basis measurements used for different sums as in Fig.~\ref{fig:ZNE-Energy}. 

\begin{figure*}[htb!]
    \centering
    \includegraphics[width=0.8\linewidth]{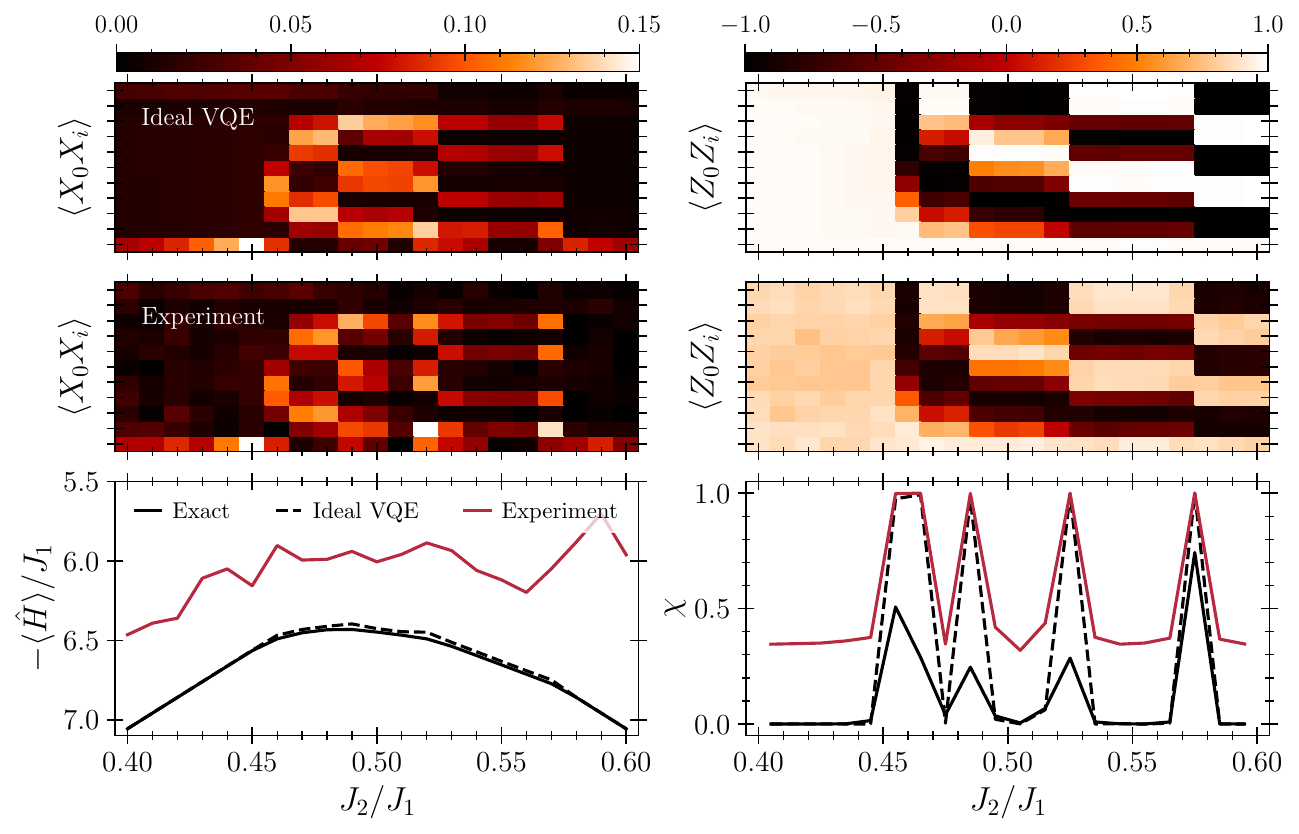}
    \vspace{-0.3cm}
    \caption{The correlation functions $\braket{X_0X_i}$ and $\braket{Z_0Z_i}$ alongside energy and FS for $N=12$ at $B_x=0.1$ with open boundary conditions. The top row shows the ideal noise-free VQE correlation functions, while the middle row shows the experimental results on \texttt{ibm\_kyoto}, starting from site $i=1$ on the $y$-axis. The $J_2/J_1$ axes are shared with Energy and FS plotted on the bottom rows. The FS is plotted at the midpoint between the two $J_2$ values used to calculate it. The differing number of peaks/transitions with respect to Figures \ref{fig:ZNE-Energy} and \ref{fig:diabat energies} is because the results plotted here are for open boundary conditions, while those are for periodic boundary conditions. All results were obtained using just TREX and Pauli twirling.}
    \label{fig: corr12 open}
\end{figure*}

Next we turn to the correlation functions $\braket{\sigma_i\sigma_j}$ for Pauli operators $\sigma$, special terms of which are present in the calculation of the energy and its derivative. In the top two panels of Fig.~\ref{fig: corr12 open} we show the ideal noise-free correlation functions $\braket{X_0X_i}$ and $\braket{Z_0Z_i}$ of the VQE calculated with a statevector simulation and the experimentally measured quantities in the second row. The differing number of phase transitions with respect to Figures \ref{fig:ZNE-Energy} and \ref{fig:diabat energies} is due to the latter results being under open boundary conditions. The corresponding energy calculation from the same data is seen in the bottom left panel. Utilizing just the simple digital modification and post processing tools of Pauli twirling and TREX, we already see that the behavior of the correlation functions is, qualitatively, quite accurate across the scanned parameters, up to slight differences in magnitude for the $\braket{Z_0Z_i}$ correlator. Due to the lower absolute magnitude of $\braket{X_0X_i}$, the fluctuations induced by noisy runs are more pronounced. Taken as a whole however, it is easy to discern similarly structured $J_2$ ranges. These qualitative trends allow one to cleanly see different `phases', i.e. ground-state configurations that the VQE takes on. 

This conclusion is further bolstered by considering the FS, shown on the bottom right panel of Fig.~\ref{fig: corr12 open}. Here we see that the noise level creates quite a high baseline within phase regions, showing that in this experiment between 35-40\% of the measured states were not the $\ket{0}^{\otimes N}$ state expected by the exact and ideal VQE simulation. However when crossing between phase regions, the measurement becomes effectively identical to the ideal VQE results. This signal to noise ratio means that the ground-state reconfiguration can be experimentally well resolved. Furthermore it is clear at a glance that the peaks in $\chi$ directly correspond to boundaries between similarly structured areas in the correlation functions corresponding to similar ground-state configurations. 

Cross-referencing these results against each other, one can also infer the reliability of outlying values in any one measure. For instance, the XX correlator at $J_2/J_1=0.51$ displays a strong drop in magnitude compared to the neighboring values. However, there is no such trend observed in the $ZZ$ correlator at this point, and the FS value at $0.505$, corresponding to the overlap between $J_2/J_1 = 0.5$ and $0.51$, does not peak, thus implying that this particular $X$ basis measurement was likely subject to some fluctuation in the underlying noise channels. 

\begin{figure*}[htb!]
    \centering
    \includegraphics[width=0.8\linewidth]{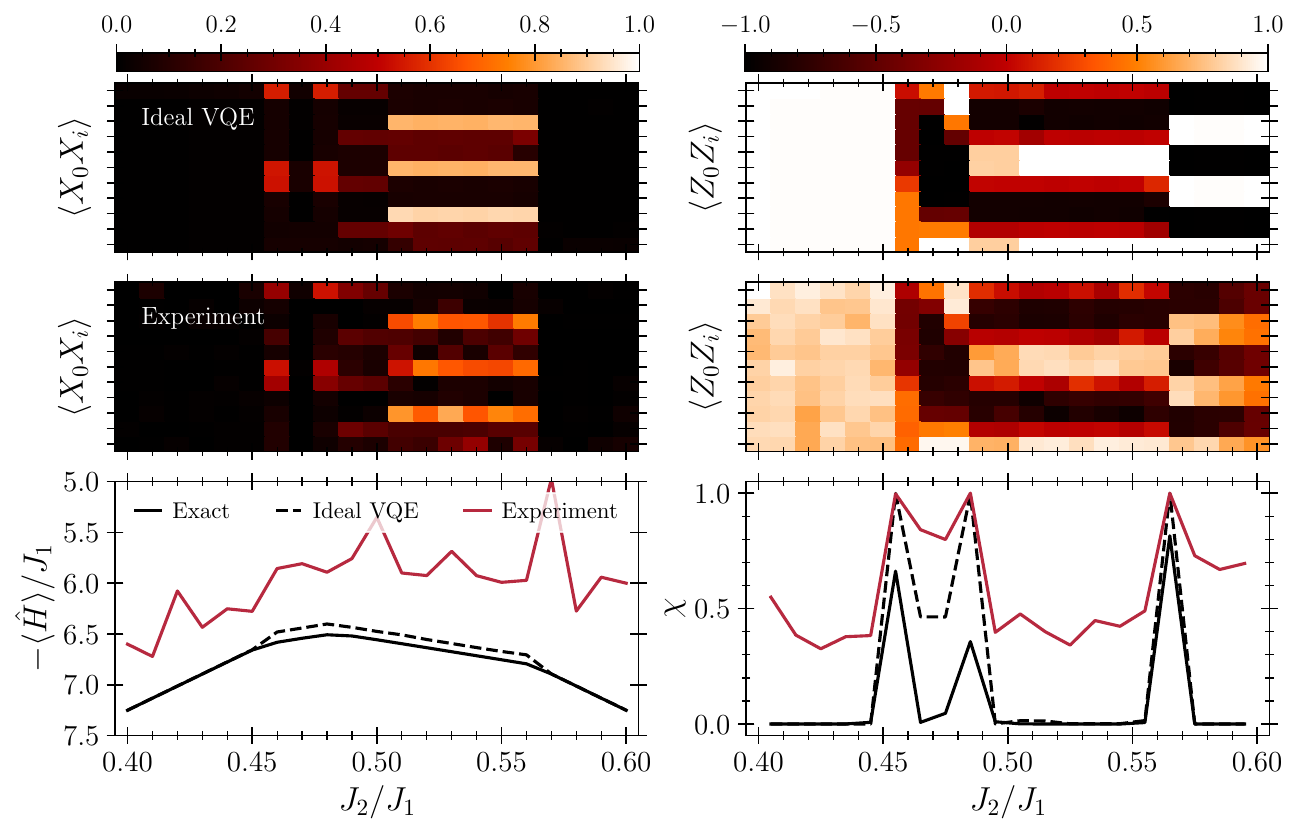}
    \vspace{-0.3cm}
    \caption{The results for $N=12,\, B_x=0.1$ under periodic boundary conditions calculated on \texttt{ibm\_osaka}. The ideal VQE results are on the first row, with experimental results on the second. Energy and fidelity susceptibility are plotted on the bottom row. The FS is plotted at the midpoint between the two $J_2$ values used to calculate it. These results were obtained with TREX.}
    \label{fig: corr12}
\end{figure*}

In Fig.~\ref{fig: corr12} we show the results of a VQE solution to the $N=12$, $B_x=0.1$ problem with periodic boundary conditions using just TREX. The heavy-hex architecture of the IBM Eagle devices allows a natural mapping to a ring of physical qubits as seen in Fig.~\ref{fig:VQE-ansatz}. In addition to the $\mathbb{Z}_2$ symmetry, this periodicity imposes a \textit{shift} symmetry. As a result of this symmetry being broken differently between neighboring $J_2$ VQE solutions the $\braket{\sigma_i\sigma_j}(p)$ matrices can be offset between $J_2$ values. Thus in plotting Fig.~\ref{fig: corr12} we align the calculated correlation function matrices between $J_2$ values as much as possible. We see that even with a substantial amount of noise-induced fluctuation in the energy calculation, the qualitative trend in the correlation functions is still quite well behaved. In these experimental results it is evident that that the signal to noise ratio in $\chi$ is higher than in Fig.~\ref{fig: corr12 open}. 

However, we can again cross reference the results to infer the `hidden' noise-free properties of the VQE solutions. In the range $J_2/J_1=(0.45,0.49)$, $\chi$ displays several large values. Furthermore, we see from the measured data that rearrangements in the $XX$ and $ZZ$ observables occur between each of these values. Thus, even if we had trained these VQE solutions blindly on a real device, we could infer that at the very least, the parameterized wavefunctions in the range has some reconfiguration, as indeed we see from the ideal VQE FS calculation in dashed black lines. Another example is the range $J_2/J_1=[0.57,0.6]$. The FS in this region is also quite high, but by looking at the correlation values it is evident that $XX$ remains near zero, while $ZZ$ retains the same $\braket{2,2}$ structure, even though the intensity fades. From this we can infer that the FS calculation data may not be reliable. 

\begin{figure*}[htb!]
    \centering
    \includegraphics[width=0.8\linewidth]{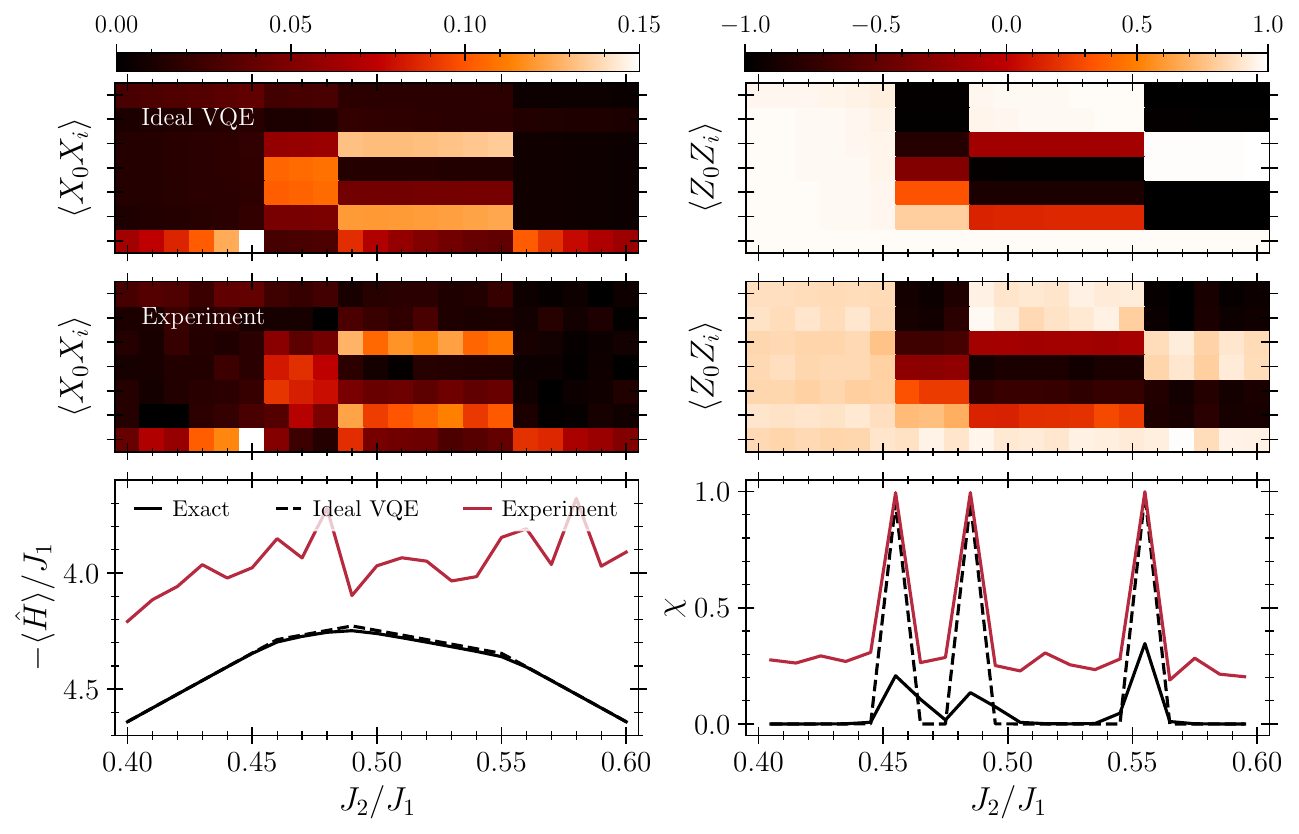}
    \vspace{-0.3cm}
    \caption{The results for $N=8,\, B_x=0.1$ with open boundary conditions calculated on \texttt{ibm\_algiers} The ideal VQE results are on the first row, with experimental results on the second.  Energy and fidelity susceptibility are plotted on the bottom row. These results were obtained with no error mitigation techniques. }
    \label{fig: corr8}
\end{figure*}

At the level of correlation functions, we found that in some cases their quantitative accuracy could be improved under TREX and twirling, although the effects are not easily discernible on a color scale plot. See figures \ref{fig:XX-correlator} and \ref{fig:XX-correlator-EM} in the Appendix to see the corrections obtained for the $XX$ correlator in N=12 under open boundary conditions. However these effects are minor enough that in figures \ref{fig: corr8} and \ref{fig:corr4} we show results for N=8 and N=4 respectively, under open boundary conditions with no EM techniques used at all. Here one sees that the qualitative trends in the correlation functions are still quite accurate, while the signal to noise ratio in FS is approximately a factor of three.

\section{Discussion and Outlook}\label{sec:Conclusion}
We found that from a practical standpoint, when using NISQ devices to calculate the energy using a VQE ansatz, even if one uses Pauli noise shaping, read out and zero-noise extrapolation error mitigation methods, the resulting value is essentially useless. Without ZNE the value is virtually guaranteed to be much larger than the ideal noise-free result. However, with ZNE it could well be that extrapolation from fits with large uncertainties render a non-physical answer below the exact ground-state energy. Conversely, when using precisely the same measurement data used in the energy calculation, we can explore the nature of the physically encoded and executed VQE wavefunction by analyzing the data through a variety of observables. For example looking at the derivative of the energy through our parameterized slice of phase space, as estimated by the Hellmann-Feynman theorem, we saw that there was very clearly phase information about the wavefunction encoded in the experimental data, which was far more robust to noise than the value of the energy. Of course this is in a qualitative sense, as the quantitative error is still non-negligible if one is thinking of the accuracy required from a quantum chemistry perspective. However, the trends between data points corresponding to differing phases are unmistakably clean.

Similarly, by focusing on `distributed' observables, i.e. the correlation function across all sites, we argue that the qualitative trend of these values gives more usable information. Since for well converged solutions there is a similar ground-state configuration between VQE wavefunctions within the same region, the underlying trends within the highly structured correlation functions of those states are easier to discern, despite the fluctuations induced at particular values due to noise. In particular for our problem where we scan through several distinct ground-state phases, the sharp difference in these quantities gives clear, noise-robust signals of ground-state rearrangement. The point is not so much that a particular correlation function value $\braket{\sigma_i\sigma_j}$ for a particular phase is numerically accurate or not, it's that the trends of these functions across the phase space provide unambiguous experimental data identifying clear regions of similar structure. Therefore we assert that using such observables to identify quantum phase transitions in VQE solutions could conceivably constitute a presently existing, noise robust utility of NISQ devices. 

Alongside this tool, we adapted the well established Fidelity Susceptibility measure to partially accommodate the limitations of the VQE ansatz, and found that it provides a high signal-to-noise measure of phase boundaries being crossed between VQE solutions. For the model studied here, this allows one to quickly cross check the spin correlation functions with the FS measure in order to determine whether there was a true ground-state rearrangement, or if there was some partial, differentiated symmetry breaking across symmetry sectors between VQE solutions. For other problems, e.g. topological phase transitions, where local order parameters do not show discontinuities between phases, the sucess of using the FS in condensed matter theory could be easily extended to NISQ computation. This provides a conceptually simple method of detecting topological phases compared to say identifying non-local string order parameters \cite{Smith2022} or training Quantum Convolutional Neural Networks \cite{Liu2023,zapletal2023errortolerant,Herrmann2022}. 

While these results have shown that one can use NISQ devices to detect phases and their boundaries in VQE solutions, it is important to emphasize that the spirit of this work is not limited to the VQE algorithm, and can indeed be extended to any method which prepares ground state solutions to particular models. Going forward, to be able to demonstrate a true \textit{advantage} over classical computation requires finding ground states for classically intractably large system sizes. In the case of the VQE, which is one of the most promising near term algorithms, this naturally requires implementing an optimization loop on real devices. Given the active research in compensating for problems of barren plateaus from overparameterization \cite{Larocca2022,Herasymenko2021} and noise \cite{Wang2021,Schumann2023}, our results suggest that characterization of complex zero temperature phase diagrams in large systems can be a real near term advantage for NISQ computation. Furthermore, while the interplay between EM and trainability of VQAs is an area of active research \cite{Wang2021_2,filippov2022matrix, Tilly2022}, fundamental limits on the ability of EM to fully compensate for noise of large system sizes have been derived \cite{Quek2022,Takagi2022}. This raises the possibility that simple measures like the ones we propose, bolstered by stable EM methods like digitally shaping the noise via twirling and post-processing methods like TREX, could still retain use in the system size vs. error threshold range between EM and fully fault tolerant computation. 

While these are interesting speculations worthy of further investigation, we end by noting a few concrete uses of phase identification in relatively smaller systems. An obvious example would be larger system sizes for the ANNNI model itself, useful for simulating the properties of the frustrated Ising magnet Ca$_3$Co$_2$O$_6$ \cite{Batista2012,Zapf2022} or other systems which display similar complicated phases which require scaling to large system sizes in order to identify \cite{Mila2022}. Furthermore, our approach dovetails well with improvements in using VQAs for resolving excited states \cite{Higgott2019} in particular in the context of conical intersections and avoided crossings \cite{Gocho2023, Omiya2022,Yalouz2022,Koridon2024hybridquantum}. If one can detect rearrangements in the electronic structure on either side of a conical intersection as encoded in physical qubit correlation functions, then there could be space for a quantum advantage over classical electronic structure methods, which already struggle at medium sized molecules. Applications could include distinguishing Jahn-Teller Effects (JTE) due to level crossings between the ground state and higher lying states such as in the fullerene C$_{60}$ molecule, versus pseudo JTE (PJTE) coming from small gapped avoided crossings such as in hemoglobin, in which the FS would in principle not pick up a ground-state rearrangement. These effects are ubiquitous in polyatomic molecular chemistry: for a comprehensive review with extensions to periodic systems see \cite{Bersuker2021}. 

In conclusion we found that by using a combination of simple target observables, we can reliably detect the properties of ground state configurations in a noise robust manner on presently existing NISQ devices by interrogation of the experimental data from multiple angles, thus making the corruption of one observable by noise less detrimental. We demonstrated this utility experimentally by detecting rearrangement between phase domains for pre-optimized VQE solutions to a frustrated spin model. Our results have the potential for direct application to circuits which prepare ground state solutions for medium and large sized systems, and could be useful even in cases where error mitigated observables are prohibitively expensive. 

\section*{Acknowledgments}

Work at Los Alamos was supported by the LANL LDRD Program, and in part by the Center for Integrated Nanotechnologies, a DOE BES user facility, in partnership with the LANL Institutional Computing Program for computational resources. We acknowledge the use of IBM Quantum services for this work. The views expressed are those of the authors, and do not reflect the official policy or position of IBM or the IBM Quantum team.

\appendix
\section{EM Improvements to ZNE}
\begin{figure}
    \centering
    \includegraphics[width=\linewidth]{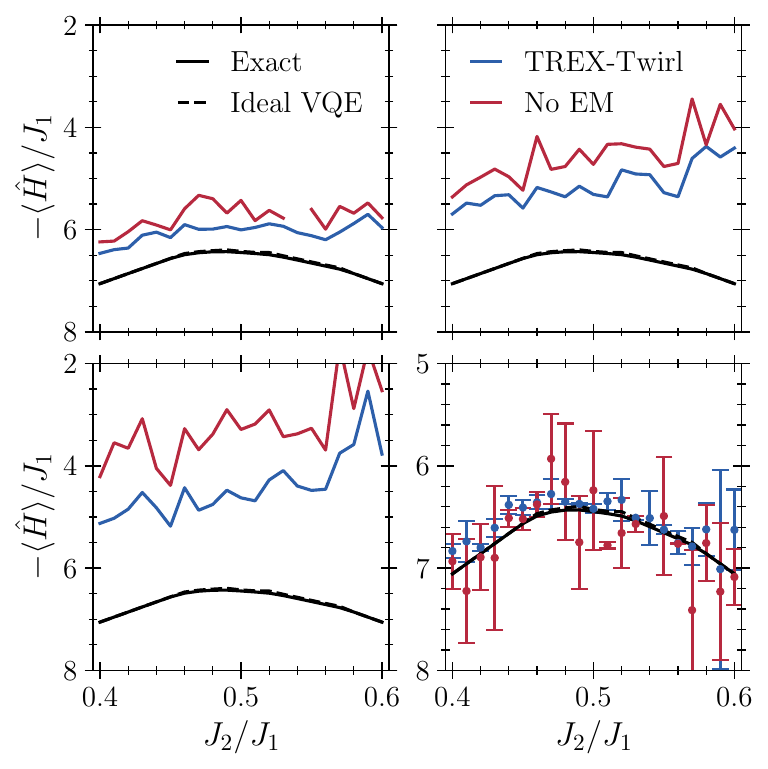}
    \vspace{-0.5cm}
    \caption{The effect of EM on noise scaling via insertion of physical application of logically redundant CX gates. These results are for $N=12,\ B_x = 0.1$ under open boundary conditions. The number of physically applied CX gates, corresponding to the $\lambda$ noise scaling, is 1, 3 and 5 in panels (a), (b) and (c), respectively. The resulting ZNE fit is shown in (d). Note that the y axis scale is different in this panel. All results here were calculated on \texttt{ibm\_kyoto}. The gap in the data in (a) is due to an unrecoverable error in the \texttt{QiskitRuntime} service for the $J_2/J_1=0.54$ run. }
    \label{fig:CX scaling}
\end{figure}
We found that mollifying the effects of readout error via TREX alongside noise shaping by Pauli twirling had a non-trivial effect on the quality of the resulting ZNE fit to the energy. In Fig.~\ref{fig:CX scaling} we show the energy calculation as it appears under increasing logically redundant but physically applied CX gates, where every CX in the ansatz is replaced with one, three or five CX cycles. In addition to the raw results having consistently higher values than those with TREX and twirling, their errors increase at a greater rate compared with the TREX-Twirl results, seen by the increasing gap between them going from $\lambda=1$ to $\lambda=5$. Subsequently the resulting fit has a larger variance, leading to larger uncertainty in the extrapolated values, alongside greater fluctuation in the extrapolated mean. In the worst cases, this leads to several expected values \emph{below} the physically bounded exact ground-state value. In contrast, while for some runs, in particular closer to the $\braket{2,2}$ region, the EM extrapolation also has high uncertainty, the expectation values are consistently closer to the ideal results and, within the error, always physical.

\section{EM Improvements to Correlation Functions}

\begin{figure}
    \centering
    \includegraphics[width=0.9\linewidth]{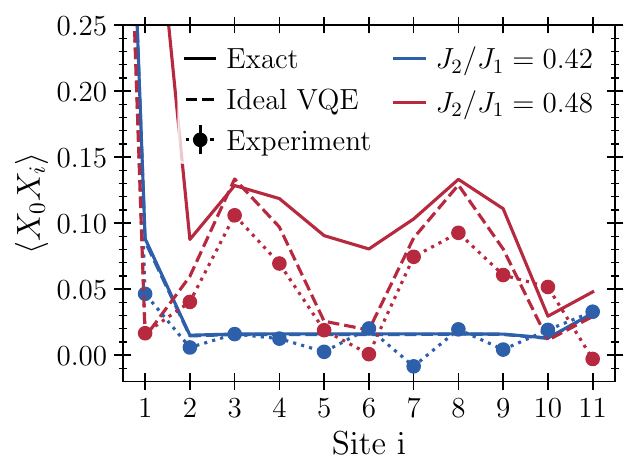}
    \vspace{-0.4cm}
    \caption{The XX correlation function for $N=12,\ B_x = 0.1$ under open boundary conditions calculated on \texttt{ibm\_osaka}, compared to exact results and the ideal VQE results. The experiments were run with no error mitigation techniques.}
    \label{fig:XX-correlator}
\end{figure}

\begin{figure}
    \centering
    \includegraphics[width=0.9\linewidth]{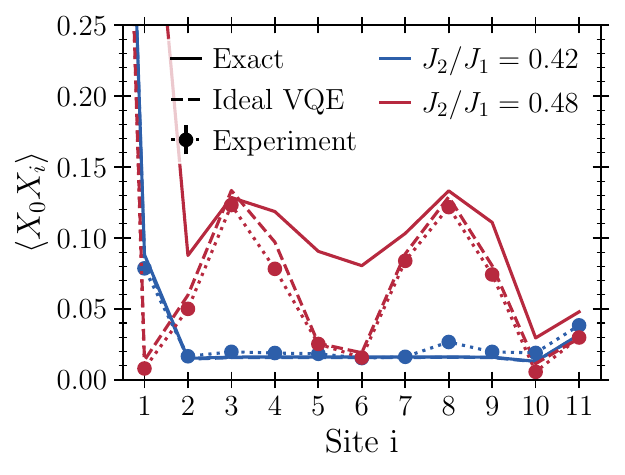}
    \vspace{-0.4cm}
    \caption{The XX correlation function for $N=12,\ B_x = 0.1$ under open boundary conditions calculated on \texttt{ibm\_osaka}, compared to exact results and the ideal VQE results. The experiments were run with TREX and Twirling.}
    \label{fig:XX-correlator-EM}
\end{figure}

While one of the primary points of this paper has been to argue that the specific numerical values of the $\braket{\sigma_i\sigma_j}$ correlation functions are not as relevant as the fact that their structured nature yields reliable signals under ground-state rearrangement, we nonetheless note that the quantitative accuracy can be further improved through the EM methods we explored here, thus improving their reliability. In Fig.~\ref{fig:XX-correlator} we show slices of the $XX$ correlator in the ferromagnetic ($J_2/J_1=0.42$) and floating ($J_2/J_1=0.48$) phases without any error mitigation applied. One sees that there is simultaneously large fluctuation in the ferromagnetic phase, while the floating phase shows strong deviation from the characteristic curve. Again, although these are clearly quantitatively quite inaccurate, they are still quite distinct with respect to one another.

Looking at Fig.~\ref{fig:XX-correlator-EM} where TREX and twirling are performed, we see that the flat line of the ferromagnetic phase and the oscillations of the floating phase are both quantitatively much better captured, in particular the amplitude of the floating phase signal becomes quite accurate with respect to to the ideal VQE results. These minor digital manipulation and post-processing routines are sufficient to quite strongly improve the signal, and the distinction between the two phases loses any ambiguity. 

\onecolumngrid

\section{N=4 data}
\begin{figure*}
    \centering
    \includegraphics[width=0.8\linewidth]{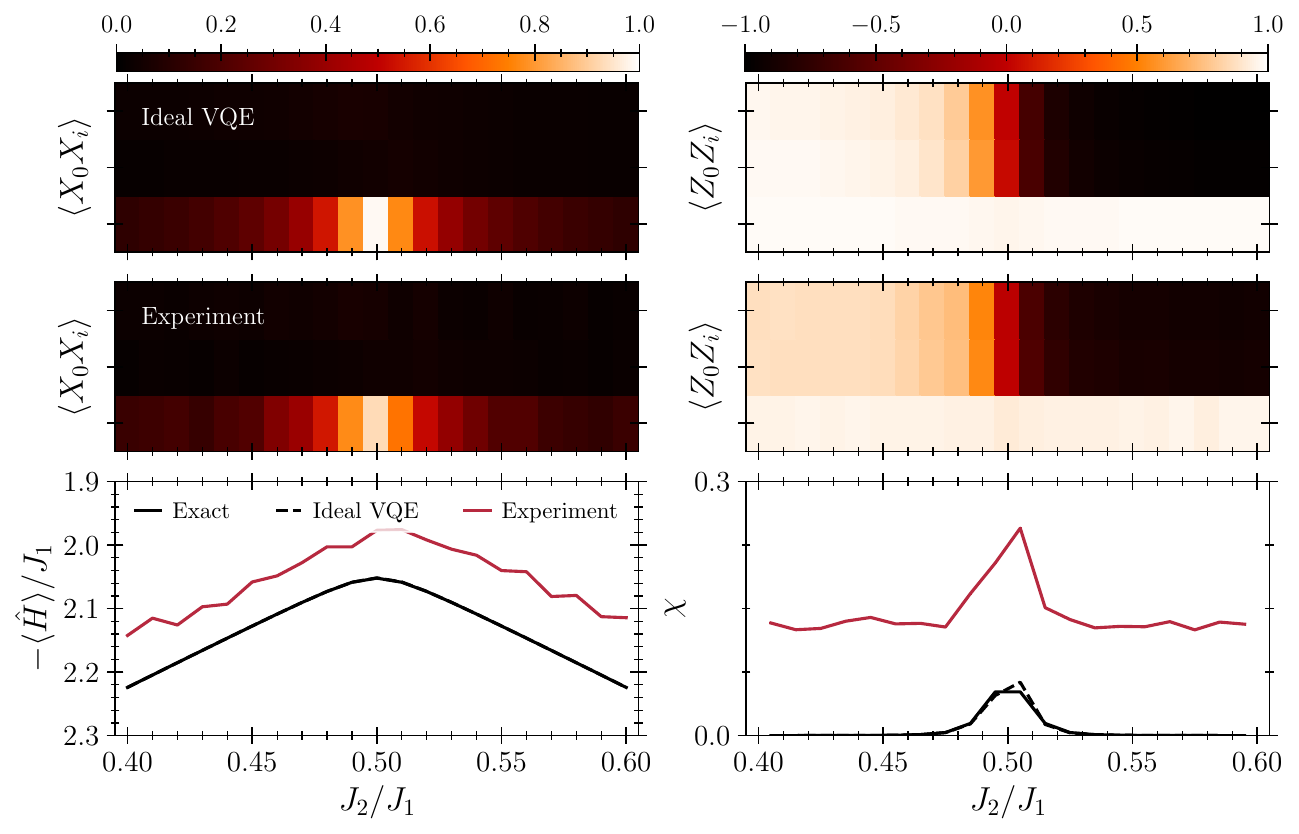}
    \vspace{-0.3cm}
    \caption{The correlation function for $N=4,\ B_x = 0.1$ under open boundary conditions. The ideal VQE results are on the first row, with experimental results on the second.  Energy and fidelity susceptibility are plotted on the bottom row. These results were obtained with no error mititgation techniques. These results were calculated on \texttt{ibm\_cusco}.}
    \label{fig:corr4}
\end{figure*}
For completeness we here include the results of an $N=4$ calculation in Fig.\,\ref{fig:corr4}, in order to demonstrate that our results hold for a variety of system sizes. In this case the correlation functions are far less structured, yet clearly show a transition across the range, while the energy calculation is also qualitatively well behaved.

\clearpage

\twocolumngrid

\bibliography{annni}

\end{document}